\documentclass[twocolumn,superscriptaddress,amsmath,amssymb]{revtex4}

\usepackage{graphicx}
\usepackage{dcolumn}
\usepackage{bm}
\usepackage{ulem}
\newcommand{\nc}{\newcommand}
\nc{\be}{\begin{eqnarray}}
\nc{\ee}{\end{eqnarray}}
\nc{\bea}{\begin{eqnarray}}
\nc{\eea}{\end{eqnarray}}
\nc{\bean}{\begin{eqnarray*}}
\nc{\eean}{\end{eqnarray*}}

\begin{document}

\title{Abrupt change in the energy gap of superconducting Ba$_{1-x}$K$_x$Fe$_2$As$_2$ single crystals with hole doping}

\author{W.~Malaeb}
\affiliation{Institute for Solid State Physics (ISSP), University of
Tokyo, Kashiwa-no-ha, Kashiwa, Chiba 277-8561, Japan}
\affiliation{TRIP, JST, Chiyoda-ku, Tokyo 102-0075, Japan}

\author{T.~Shimojima}
\affiliation{Institute for Solid State Physics (ISSP), University of
Tokyo, Kashiwa-no-ha, Kashiwa, Chiba 277-8561, Japan}
\affiliation{CREST, JST, Chiyoda-ku, Tokyo 102-0075, Japan}

\author{Y.~Ishida}
\affiliation{Institute for Solid State Physics (ISSP), University of
Tokyo, Kashiwa-no-ha, Kashiwa, Chiba 277-8561, Japan}
\affiliation{CREST, JST, Chiyoda-ku, Tokyo 102-0075, Japan}

\author{K.~Okazaki}
\affiliation{Institute for Solid State Physics (ISSP), University of
Tokyo, Kashiwa-no-ha, Kashiwa, Chiba 277-8561, Japan}
\affiliation{CREST, JST, Chiyoda-ku, Tokyo 102-0075, Japan}

\author{Y.~Ota}
\affiliation{Institute for Solid State Physics (ISSP), University of
Tokyo, Kashiwa-no-ha, Kashiwa, Chiba 277-8561, Japan}


\author{K.~Ohgushi}
\affiliation{Institute for Solid State Physics (ISSP), University of
Tokyo, Kashiwa-no-ha, Kashiwa, Chiba 277-8561, Japan}
\affiliation{TRIP, JST, Chiyoda-ku, Tokyo 102-0075, Japan}

\author{K.~Kihou}
\affiliation{TRIP, JST, Chiyoda-ku, Tokyo 102-0075, Japan}
\affiliation{National Institute of Advanced Industrial Science and
Technology (AIST), Tsukuba, Ibaraki 305-8568, Japan}

\author{T.~Saito}
\affiliation{Department of Physics, Chiba University, Chiba
263-8522, Japan}

\author{C.H.~Lee}
\affiliation{TRIP, JST, Chiyoda-ku, Tokyo 102-0075, Japan}
\affiliation{National Institute of Advanced Industrial Science and
Technology (AIST), Tsukuba, Ibaraki 305-8568, Japan}

\author{S.~Ishida}
\affiliation{National Institute of Advanced Industrial Science and
Technology (AIST), Tsukuba, Ibaraki 305-8568, Japan}
\affiliation{Department of Physics, University of Tokyo, Tokyo
113-8656, Japan}

\author{M.~Nakajima}
\affiliation{National Institute of Advanced Industrial Science and
Technology (AIST), Tsukuba, Ibaraki 305-8568, Japan}
\affiliation{Department of Physics, University of Tokyo, Tokyo
113-8656, Japan}

\author{S.~Uchida}
\affiliation{Department of Physics, University of Tokyo, Tokyo
113-8656, Japan}

\author{H.~Fukazawa}
\affiliation{TRIP, JST, Chiyoda-ku, Tokyo 102-0075, Japan}
\affiliation{Department of Physics, Chiba University, Chiba
263-8522, Japan}

\author{Y.~Kohori}
\affiliation{TRIP, JST, Chiyoda-ku, Tokyo 102-0075, Japan}
\affiliation{Department of Physics, Chiba University, Chiba
263-8522, Japan}

\author{A.~Iyo}
\affiliation{TRIP, JST, Chiyoda-ku, Tokyo 102-0075, Japan}
\affiliation{National Institute of Advanced Industrial Science and
Technology (AIST), Tsukuba, Ibaraki 305-8568, Japan}

\author{H.~Eisaki}
\affiliation{TRIP, JST, Chiyoda-ku, Tokyo 102-0075, Japan}
\affiliation{National Institute of Advanced Industrial Science and
Technology (AIST), Tsukuba, Ibaraki 305-8568, Japan}

\author{C.-T.~Chen}
\affiliation{Beijing Center for Crystal R and D, Chinese Academy of
Science (CAS), Zhongguancun, Beijing 100190, China}

\author{S.~Watanabe}
\affiliation{Institute for Solid State Physics (ISSP), University of
Tokyo, Kashiwa-no-ha, Kashiwa, Chiba 277-8561, Japan}

\author{H.~Ikeda}
\affiliation{Department of Physics, Kyoto University, Sakyo-ku,
Kyoto 606-8502, Japan}

\author{S.~Shin}
\affiliation{Institute for Solid State Physics (ISSP), University of
Tokyo, Kashiwa-no-ha, Kashiwa, Chiba 277-8561, Japan}
\affiliation{TRIP, JST, Chiyoda-ku, Tokyo 102-0075, Japan}
\affiliation{CREST, JST, Chiyoda-ku, Tokyo 102-0075, Japan}

\date{\today}


\begin{abstract}

We performed a Laser angle-resolved photoemission spectroscopy
(ARPES) study on a wide doping range of Ba$_{1-x}$K$_x$Fe$_2$As$_2$
(BaK) and precisely determined the doping evolution of the
superconducting (SC) gaps in this compound. The gap size of the
outer hole Fermi surface (FS) sheet around the Brillioun zone (BZ)
center shows an abrupt drop with overdoping (for $x \gtrsim$ 0.6)
while the inner and middle FS gaps roughly scale with $T_c$. This is
accompanied by the simultaneous disappearance of the electron FS
sheet with similar orbital character at the BZ corner. These results
browse the different contributions of $X^2-Y^2$ and $XZ/YZ$ orbitals
to superconductivity in BaK and can be hardly completely reproduced
by the available theories on iron-based superconductors.

\end{abstract}

\maketitle



A detailed knowledge of the superconducting (SC) gap size and
anisotropy is essential for understanding the electron pairing
mechanism in iron pnictides \cite{Kamihara_JACS08}. From the early
stage, it was thought that electron pairing is likely to be of
magnetic origin induced by antiferromagnetic (AFM) spin-fluctuations
(SF) between disconnected Fermi surface (FS) sheets
\cite{Mazin_PRL08, Kuroki_PRL08, Kuroki_PRB09}. Within this
inter-band scattering picture, a full sign-reversing superconducting
(SC) gap ($s^{+-}$) is expected which was supported by the findings
of angle-resolved photoemission spectroscopy (ARPES) studies
\cite{Ding, Borisenko_PRB09, Nakayama, Borisenko_arXiv11} and other
experimental results \cite{Hanaguri_Science10, Christianson_Nat08,
Hiraishi_JPSJ09}. However, some theoretical studies
\cite{Kontani_PRL, Kontani_PRB10, Yanagi_JPSJ10} have pointed out to
a possible role of the orbital degrees of freedom in the pairing
mechanism. This was supported by a recent Laser ARPES study on SC
BaFe$_2$(As$_{1-x}$P$_x$)$_2$ (AsP) and optimally-doped (OP)
Ba$_{1-x}$K$_x$Fe$_2$As$_2$ (BaK) compounds
\cite{Shimojima_Science11} which found nearly orbital-independent SC
gaps on the hole-like FS sheets around the Brillouin Zone (BZ)
center. Nevertheless, there is an increasing evidence that the SC
gap in iron pnictides is not universal \cite{Kuroki_JPSJ11}, as
there is evidence for gap nodes in some compounds such as in AsP
\cite{Kuroki_JPSJ11, Hashimoto1_PRB10, AsP}, KFe$_2$As$_2$
\cite{K122, Hashimoto2_PRB10, Kuroki_K122} and
Ba(Fe$_{1-x}$Co$_x$)$_2$As$_2$ \cite{FeCo}.

Among other iron pnictides, BaK is special regarding the persistence
of superconductivity in this compound up to the end member
KFe$_2$As$_2$. Also the SC gaps in BaK are expected to change with
doping from fully-opened gaps near the OP region \cite{Ding,
Borisenko_PRB09, Nakayama} to nodal gaps at the end member
KFe$_2$As$_2$ \cite{K122, Hashimoto2_PRB10, Kuroki_K122}. Therefore,
BaK is an ideal system to study the doping evolution of SC gaps.
Moreover, it is important to investigate the significance of each
orbital component to the unconventional pairing mechanism in BaK and
its doping evolution which will be a crucial test for the validity
of available theories on iron pnictides. While ARPES is the most
direct experimental technique in probing the electronic structure of
solids and addressing the above issues, there is still little
consensus on the SC gaps of iron pnictides, including BaK,
determined from ARPES. Indeed, part of this originates from the
limited experimental resolution. Here, we have implemented Laser
ARPES with high energy resolution and bulk sensitivity on a wide
doping range of BaK extending from the underdoped (UD) to the
overdoped (OD) regime. This enabled the assignment of the orbital
character of each of the hole bands around the BZ center and the
precise determination of the doping evolution of the SC gaps on each
of these bands.



High-quality single crystals of underdoped (UD) ($x \sim$ 0.2, 0.3)
and overdoped (OD) ($x \sim$ 0.5, 0.6, 0.7 and 0.85)
Ba$_{1-x}$K$_x$Fe$_2$As$_2$ (BaK) were grown by using the FeAs
\cite{Ohgushi} and KAs \cite{Kihou} flux respectively. Laser ARPES
measurements were carried out at ISSP, University of Tokyo. We used
a VG-Scienta R4000 as an electron analyzer and a VUV Laser of $h\nu$
= 6.994 eV as a light source. In this spectrometer \cite{Kiss},
changing the polarization vector of the laser source is possible by
rotating the half-wave ($\lambda$/2) plate without changing the
optical path. The Fermi level ($E_F$) calibration of the samples was
done by referring to that of gold and the energy resolution was set
to $\sim$3 meV. The samples were cleaved at a temperature $T \sim$
30 K or less in an ultra-high vacuum of less than 5 $\times$
10$^{-11}$ Torr. Other ARPES measurements were performed using a
Helium discharge source with $h\nu$ = 21.218 eV. In this case, the
electron analyzer is similar to that of the Laser spectrometer. Also
the sample cleaving and $E_F$ calibration are similar to the Laser
ARPES case, however the energy resolution in this case was set to
$\sim$10 meV.



\begin{figure}[htb]
\begin{center}
\includegraphics[width=8.5cm]{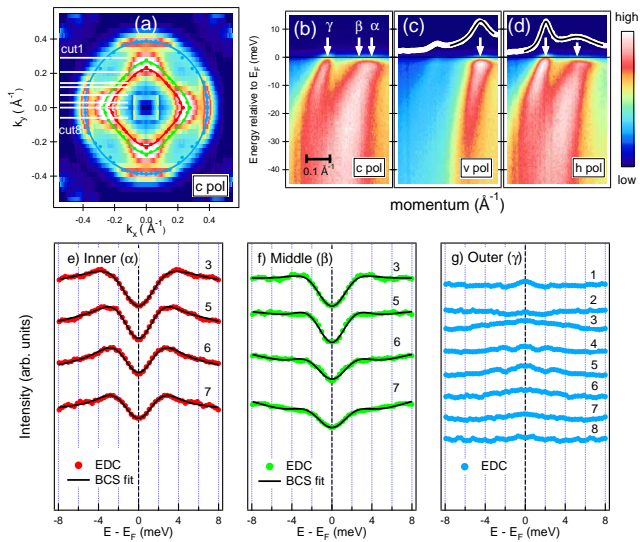}
\caption{\label{fig1} (color online) Laser ARPES data of heavily OD
BaK ($x \sim$ 0.7, $T_c \sim$ 20 K): (a) FS map in the $k_x$-$k_y$
plane measured at $T \sim$ 7 K with circularly-polarized light. The
intensity was integrated within a window of $\pm$5 meV around $E_F$
and four-fold symmetrization was done. The dots indicate the $k_F$
points determined from MDCs fitting and the colored curves are
guides to the eye. The cuts are labeled by white lines. (b-d) ARPES
$E$-$k$ intensity maps showing the band dispersions of BaK ($x \sim$
0.7) at $T \sim$ 7 K taken along cut5 in panel (a) using different
polarization directions of the incident Laser ($c$, $v$ and $h$).
The corresponding MDC (white curves) are shown in top of panels (c)
and (d) with their Lorentzian fitting (solid black curves). The
white arrows represent the $k_F$ points determined from the MDC peak
positions. (e-g) The symmetrized EDCs corresponding to several cuts
taken on the FS contour (panel a). (e), (f) and (g) correspond to
the inner $\alpha$, middle $\beta$ and outer $\gamma$ bands
respectively. The black solid lines represent the EDCs fitting to
the BCS function (refer to text). }
\end{center}
\end{figure}

The FS map of heavily OD BaK ($x \sim$ 0.7) taken at low temperature
($\sim$7 K) is displayed in Fig. \ref{fig1}(a). Three FS sheets
could be resolved around the BZ center as expected by band-structure
calculations \cite{BandStructure1, BandStructure2}. This is
confirmed by the polarization-dependent $E$-$k$ plots and the
corresponding momentum distribution curves (MDCs) at $E_F$ shown in
panels (b-d) which correspond to cut5 in panel (a). Only the inner
hole-like band ($\alpha$) appears clearly with $v$-polarization
[panel (c)] while the middle ($\beta$) and outer ($\gamma$) bands
are better observed with $h$-polarization [panel (d)]. This is an
experimental demonstration of the multi-orbital character of the
near-$E_F$ bands in BaK. We have carried out a detailed
polarization-dependent study \cite{Supp} and by following the
selection rules \cite{Damascelli, Feng}, we could assign the orbital
character of the hole bands around the BZ center as follows: The
outer band has mainly $X^2-Y^2$ orbital character, the middle band
has $XZ$/$YZ$(+$Z^2$) orbital character and the inner band is mainly
characterized with $XZ$/$YZ$ character. $X$ and $Y$ refer to the
tetragonal unit cell axes oriented 45 degrees with respect to the
Fe-Fe bond direction and $Z$ is normal to the $XY$ plane.

The energy distribution curves (EDCs) corresponding to the cuts
shown in panel (a) and representing the inner, middle and outer
bands are respectively displayed in panels (e-g) of Fig.\
\ref{fig1}. These EDCs, as those in the rest of the paper,
correspond to a thin momentum window integrated around $k_F$ points.
A common method to check the opening of the SC gap is the
symmetrization of the EDCs \cite{Norman} as we have done here. With
symmetrization, it became clear that the SC gap opens for all the
cuts taken on the inner and middle FS sheets, where a dip is
observed (panels e and f). However, the symmetrized EDCs of the
outer FS sheet show a peak (panel g) implying that the SC gap size
is negligibly small for all the cuts taken here (panel a). Actually,
this is already clear from the $E$-$k$ plots of Fig.\ \ref{fig1}
where the inner and middle bands show bending, a typical indication
of the SC gap opening while no similar features can be observed in
the outer band which crosses $E_F$. Moreover, we have extracted the
SC gap size precisely on the inner and middle FS sheets by fitting
their EDCs with a BCS spectral function (similar to ref
\cite{Shimojima_Science11}) and the fitting curves are shown by
black solid lines.

\begin{figure}[htb]
\begin{center}
\includegraphics[width=8cm]{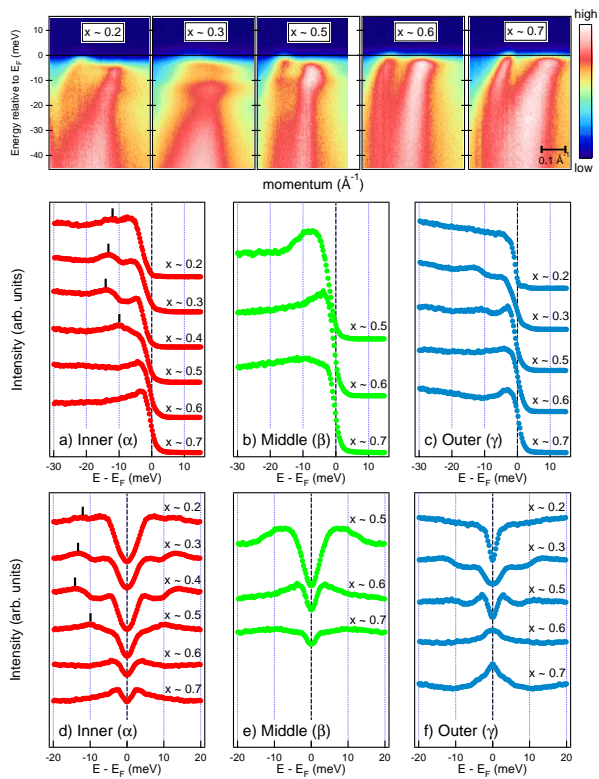}
\caption{\label{fig2} (color online) Laser ARPES data of BaK with
several dopings: $E$-$k$ plots taken at a cut along or close to the
high-symmetry line in momentum space measured in the SC state at low
temperature ($T \sim$ 7K) with circularly-polarized light. The
energy distribution curves (EDCs) corresponding to these $E$-$k$
plots where (a), (b) and (c) display the inner, middle and outer
band EDCs respectively and their symmetrized counterparts are
displayed in panels (d), (e) and (f). The black bars indicate the
position of an additional peak observed at higher binding energies
compared to the SC peak. The x=0.4 data were taken from ref.
\cite{Shimojima_Science11}. }
\end{center}
\end{figure}

\begin{figure}[htb]
\begin{center}
\includegraphics[width=7cm]{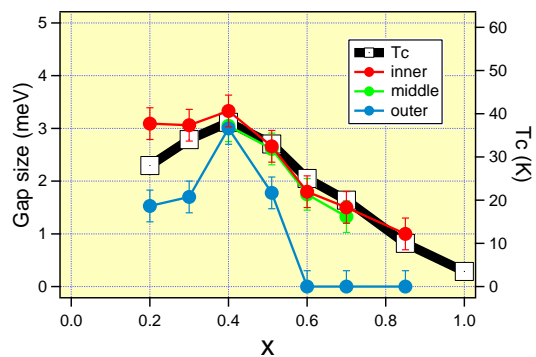}
\caption{\label{fig3} (color online) Doping dependence of the SC gap
size (left y-axis) in BaK compound on the different hole-like bands
around the BZ center determined from the data displayed in Fig.\
\ref{fig2} and the SC transition temperature $T_c$ (right y-axis).
The x=0.4 data were taken from ref. \cite{Shimojima_Science11}.}
\end{center}
\end{figure}

In order to get a deeper insight into the evolution of the SC gap
size and symmetry in BaK compound, we show in Fig.\ \ref{fig2} the
ARPES data of samples with different doping levels both in the UD
and OD regimes. The upper panel displays the $E$-$k$ plots
corresponding to a cut taken along or close to the high-symmetry
line (refer to Fig.\ \ref{fig1}) and measured in the SC state at $T
\sim$ 7 K where the SC gap clearly opens. The EDCs corresponding to
these $E$-$k$ plots are shown in panels a, b and c for the inner,
middle and outer bands respectively while their symmetrized
counterparts are displayed in panels d, e and f. The symmetrized
EDCs suggest that, for all dopings, the SC gap opens on all the
bands by showing a dip, except on the outer bands of the heavily OD
samples ($x \sim$ 0.6 and 0.7) where a peak is observed (panel f)
implying a negligibly small SC gap in this case (practically zero
within error bars). We also note that there is a non-dispersive
feature that appears in the EDCs as a second peak at high-binding
energy ($\sim$ 10 - 14 meV) as shown by black bars in panels a and d
of Fig.\ \ref{fig2}. The intensity of this peak is more enhanced in
the OP and UD regions compared to the OD region where it quickly
disappears and the band dispersion changes to a simple Bogoliubov
quasiparticle dispersion. Moreover, we determined the SC gap size
precisely by fitting the EDCs to a BCS spectral function and the
results are displayed in Fig.\ \ref{fig3}. It should be noted that
although previous ARPES studies on OP BaK \cite{Ding,
Borisenko_arXiv11} have shown signatures of the two-peak feature
observed in our data (Fig.\ \ref{fig2}), they have poorly resolved
the SC peak from the higher binding-energy peak and thus
overestimated the SC gap size on the hole FS sheets.

\begin{figure}[htb]
\begin{center}
\includegraphics[width=8.75cm]{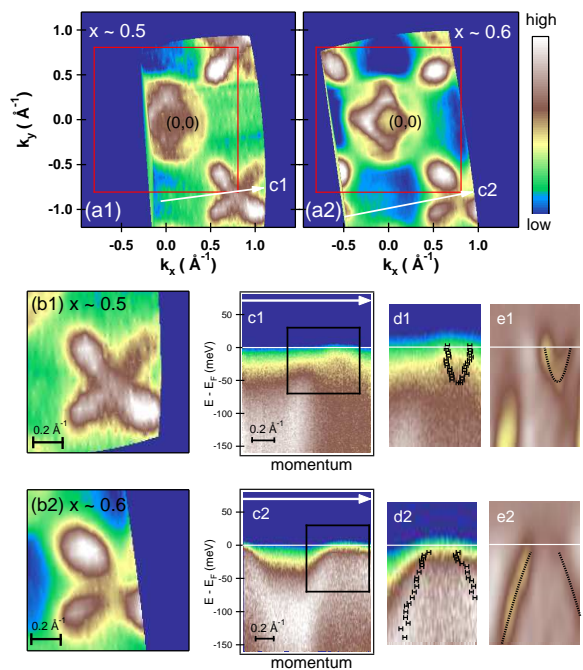}
\caption{\label{fig4} (color online) FS maps of BaK with $x \sim$
0.5 (a1) and 0.6 (a2) together with the corresponding magnified
images around the X point in panels b1 and b2. These maps were taken
by He-discharge source with $h\nu$ = 21.218 eV at $T \sim$ 7 K. The
red solid lines enclose the first Brillouin zone. (c1, c2) $E$-$k$
plots corresponding to cuts taken near the X point and represented
by white arrows (c1 at $T \sim$ 35 K for $x \sim$ 0.5 and c2 at $T
\sim$ 7 K for $x \sim$ 0.6). Magnified $E$-$k$ plots and MDC second
derivative plots corresponding to the windows enclosed by black
lines are respectively shown in panels d1, e1 for cut c1 ($x \sim$
0.5) and in panels d2, e2 for cut c2 ($x \sim$ 0.6). The dots
indicate the MDC peak positions determined from Lorentzian fitting
and the black dotted lines are guides to the eye.}
\end{center}
\end{figure}

It becomes clear from Fig.\ \ref{fig3} that by moving away from the
OP region, the orbital-independent SC gaps have become strongly
orbital dependent. The most striking feature is the abrupt drop in
the outer FS sheet gap size in the heavily OD region (for $x
\gtrsim$ 0.6) while the inner and middle FS gaps roughly scale with
$T_c$. Indeed, this is accompanied by an abrupt change in the FS
topology at $x \sim$ 0.6 as we show in Fig. \ref{fig4}. Using a He
discharge source with $h\nu$ = 21.218 eV we show the FS maps of $x
\sim$ 0.5 and 0.6 in panels a1 and a2 of Fig. \ref{fig4}
respectively with the corresponding magnified images around the X
point in panels b1 and b2. The cuts taken near the X point show that
while a small electron pocket still exists around the X point in $x
\sim$ 0.5 (cut c1) similar to OP BaK \cite{Borisenko_Nat09}, only
hole-like bands reach $E_F$ and form propeller-like FS in $x \sim$
0.6 (cut c2). This can be clearly observed in the magnified $E$-$k$
plots and their corresponding MDC second derivative plots
respectively displayed in panels d1, e1 for cut c1 and panels d2, e2
for cut c2 where an electron band clearly crosses $E_F$ near the X
point for $x \sim$ 0.5 but not for $x \sim$ 0.6. Also, this is
consistent with the observation of an intensity at the X point in
panels a1 and b1 ($x \sim$ 0.5) but not in panels a2 and b2 ($x
\sim$ 0.6). Therefore, it can be concluded that the collapse of the
outer FS gap to negligibly-small values for $x \gtrsim$ 0.6 is
probably a consequence of the disappearance of the counter electron
pocket at the X point. Such an interpretation is supported by the
fact that this gap survives up to $x \sim$ 0.5 (Fig. \ref{fig3})
where the electron pockets at the BZ corner still exist.

\begin{figure}[htb]
\begin{center}
\includegraphics[width=9.0cm]{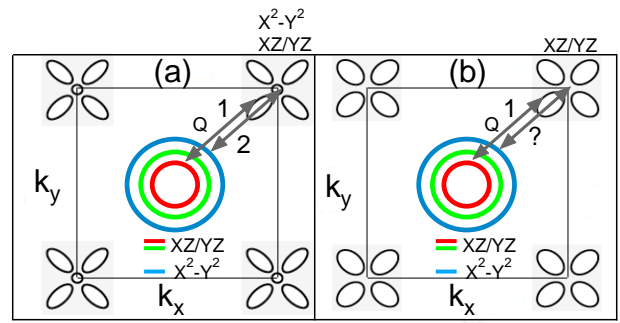}
\caption{\label{fig5} (color online) Schematics of BaK FSs with the
possible \textit{intra-orbital} interaction channels 1 and 2 between
the disconnected sheets through the wave vector $Q$ corresponding to
BaK with $x \lesssim$ 0.5 (a) and $x \gtrsim$ 0.6 (b).}
\end{center}
\end{figure}

In an attempt to understand this abrupt change in the gap structure,
we illustrate in Fig.\ \ref{fig5} the pair scattering process within
portions of the FS having similar orbital character
(\textit{intra-orbital} interactions). While both $X^2-Y^2$ and
$XZ$/$YZ$ orbital components exist at the X point around OP BaK
(Fig.\ \ref{fig5}(a)), $XZ$/$YZ$ rather than $X^2-Y^2$ orbitals has
dominant contribution at this point for the heavily OD case (Fig.\
\ref{fig5}(b)) as expected by the density functional theory (DFT)
calculations \cite{Supp}. This is consistent with the disappearance
of the electron pockets with dominant $X^2-Y^2$ orbital character
for $x \gtrsim$ 0.6 as we observed using He-discharge ARPES. On the
other hand, the SC gap on the inner and middle hole FS sheets, with
dominant $XZ$/$YZ$ orbital character, has finite values for all
doping levels investigated here and roughly scales with $T_c$ (Fig.\
\ref{fig3}). This may be a consequence of the survival of the
inter-band pair scattering related to these sheets through channel
(1) (Fig.\ \ref{fig5}). As for the UD region down to $x \sim$ 0.2,
the SC gap size at the outer sheet has finite values, unlike the
heavily OD case, because the electron pockets exit at the X point
and the \textit{intra-orbital} interactions conditions are
satisfied.

Our results browse the importance of \textit{intra-orbital}
interactions for superconductivity in BaK compound and highlight the
different roles of $X^2-Y^2$ and $XZ$/$YZ$ orbitals as illustrated
in Fig.\ \ref{fig5}. Obviously, we observe a diminishing role of the
$X^2-Y^2$ compared to the $XZ$/$YZ$ orbital components in the OD
region of BaK. This is in contrast with the OP case where
orbital-independent SC gaps were observed
\cite{Shimojima_Science11}. Indeed, the recent INS results on BaK
\cite{Castellani_arXiv1106_0071} are in agreement with our ARPES
results. The magnetic excitation observed by INS is dominated by
contributions from the $X^2-Y^2$ orbital component
\cite{Tohyama_PRB11} and its collapse in heavily OD BaK indicate the
diminishing role of $X^2-Y^2$ in this doping regime. Moreover, such
an abrupt change at the OD side is expected to have clear
implications on the thermodynamic properties, specifically the
thermal conductivity and specific heat in OD BaK compared to the OP
case.

The robustness of the outer FS negligible gap with increasing the
doping for $x \gtrsim$ 0.6 suggests that what we observe here is
different from the accidental nodes expected by theoretical studies
based on the spin-fluctuation (SF) scenario \cite{Kuroki_PRB09,
Kuroki_JPSJ11, Kuroki_K122}. Moreover, such a collapse in the outer
FS gap may not be easily reproduced by the Random Phase
Approximation (RPA) calculations. Therefore, it may be that other
factors like electron correlations effect should be included in
these calculations or theoretical calculations beyond the RPA should
be considered to account for our experimental results. Also, it may
be thought that the similar SC gap size on the inner and middle FS
sheets with dominant $XZ$/$YZ$ orbital character is remnant of the
contribution of orbital fluctuations (OF) which are believed to be
active in the OP region of BaK \cite{Shimojima_Science11}, but
obviously the OF has no contribution to the outer FS sheet in the
heavily OD region otherwise its gap would not have collapsed to
negligible values. Irrespective of the interpretation, our results
show new aspects of the SC gap of BaK compared to previous ARPES
reports \cite{Nakayama} and can be hardly completely explained by
the available theories of iron pnictides which remain to clarify the
degree of contributions of SF and/or OF in future studies.

In summary, we have precisely determined the doping evolution of the
SC gaps on the hole FS sheets around the BZ center in BaK compound
after assigning the orbital character of each of these sheets.
Unlike the inner and middle FS gaps which roughly scale with $T_c$,
the outer hole FS gap shows an abrupt drop with overdoping (for $x
\gtrsim$ 0.6) accompanied by the simultaneous disappearance of the
electron FS sheet with similar orbital character at the BZ corner.
These results browse the different contributions of $X^2-Y^2$ and
$XZ/YZ$ orbitals to superconductivity in BaK and can be hardly
completely reproduced by the available theories on iron-based
superconductors.



\section*{Acknowledgements} We thank Y.~Matsuda,
T.~Shibauchi, S.~Kasahara, K.~Hashimoto, K.~Kontani, K.~Kuroki,
R.~Arita, S.~Shamoto, M.~Ishikado, T.~Kondo, T.~Yoshida, S.~Ideta
and A.~Fujimori for informative discussions and Y.~Kotani,
K.~Koizumi and T.~Yamamoto for experimental support. This research
is supported by the Japan Society for the Promotion of Science
(JSPS) through its Funding Program for World-Leading Innovation R
and D on Science and Technology (FIRST) program.\newpage

\newpage



\newpage
\newpage
\newpage

\section*{Supplemental material}

\section{Orbital character of the hole-like bands}

In order to determine the orbital character of the hole-like bands
around the BZ center, we have considered two different geometries:
cut (A) and cut (B) rotated by 45 degrees w.r.t each others as shown
in Fig.~S1. For cut (A): $X^2-Y^2$, $Z^2$ and $XZ$ orbitals have
even parities w.r.t the mirror plane while $XY$ and $YZ$ orbitals
have odd parities. For cut (B): $X^2-Y^2$ has odd parity while $Z^2$
and $XY$ orbitals have even parities. As a result, we conclude the
following: $X^2-Y^2$ has even parity w.r.t cut (A) and odd parity
w.r.t cut (B). $XZ$/$YZ$(+$Z^2$) has even parity w.r.t both cuts (A)
and (B). $XZ$/$YZ$ has odd parity w.r.t both cuts (A) and (B). For
both cases (cuts (A) and (B)) we used two polarization directions of
the incident light: $h$ and $v$, which respectively correspond to
the horizontal and vertical directions as indicated in Fig.~S1. The
$h$-polarized light has only an even component w.r.t the mirror
plane, however both odd and even components are included in the
$v$-polarized light.

\begin{figure}[htb]
\begin{center}
\includegraphics[width=7.5cm]{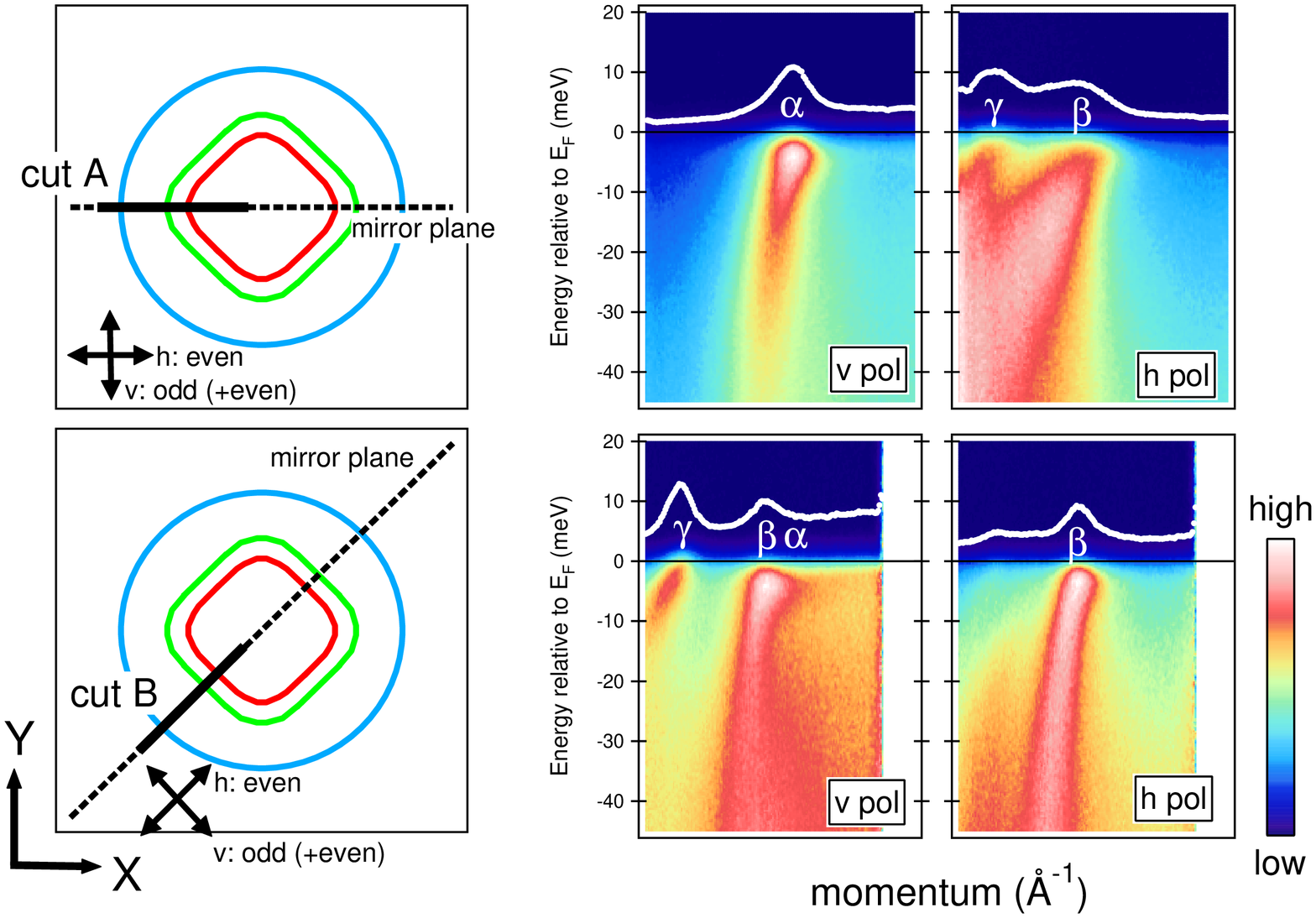}
\begin{flushleft} {\label{FigS1} FIG.~S1: Polarization-dependent Laser ARPES data of
BaK with $x \sim$ 0.6: The $E$-$k$ plots corresponding to cuts A and
B taken at different polarization directions are displayed in the
upper and lower panels respectively. The inner, middle and outer
bands are respectively labeled $\alpha$, $\beta$ and $\gamma$. The
white curves represent the corresponding MDCs at $E_F$. The
directions of cuts A and B (black solid lines) which also represent
the analyzer slit direction are illustrated in the left-hand side
panels together with the FS contours and the polarization direction
of the incident light. }
\end{flushleft}
\end{center}
\end{figure}

In cut (A), the outer hole band could be clearly observed using
$h$-polarized rather than $v$-polarized light. However, in cut (B)
the opposite is true. Therefore, the dominant orbital component of
the outer hole band is $X^2-Y^2$. The middle band could be observed
for both cuts (A) and (B) using $h$-polarized light which implies
that it has $XZ$/$YZ$(+$Z^2$) orbital character. Finally the inner
band could be observed for both cuts (A) and (B) using $v$-polarized
light which implies that it has $XZ$/$YZ$ orbital character.

\section{Dominant contributions of $X^2-Y^2$ and $XZ$/$YZ$ orbitals
expected from DFT calculations}

We performed density functional theory (DFT) calculations to compare
the orbital character of the FS sheets at the X point in the UD and
OD regions of BaK. The results are displayed in Fig.~S2 which show
three-dimensional FSs of KFe$_2$As$_2$ (c, d) and UD BaK (a, b)
obtained by shifting $E_F$ towards higher positions due to the
reduction in hole doping. It should be noted that the FSs of heavily
OD BaK are similar to those of KFe$_2$As$_2$. These plots were
obtained by using the Wien2k code \cite{Wien2k_code} with
experimental lattice parameters \cite{Rotter}. Color on the FS
sheets depict orbital weights calculated by the Wannier90
\cite{Mostofi} via the Wien2Wannier interface \cite{Kune}.

\begin{figure}[htb]
\begin{center}
\includegraphics[width=8cm]{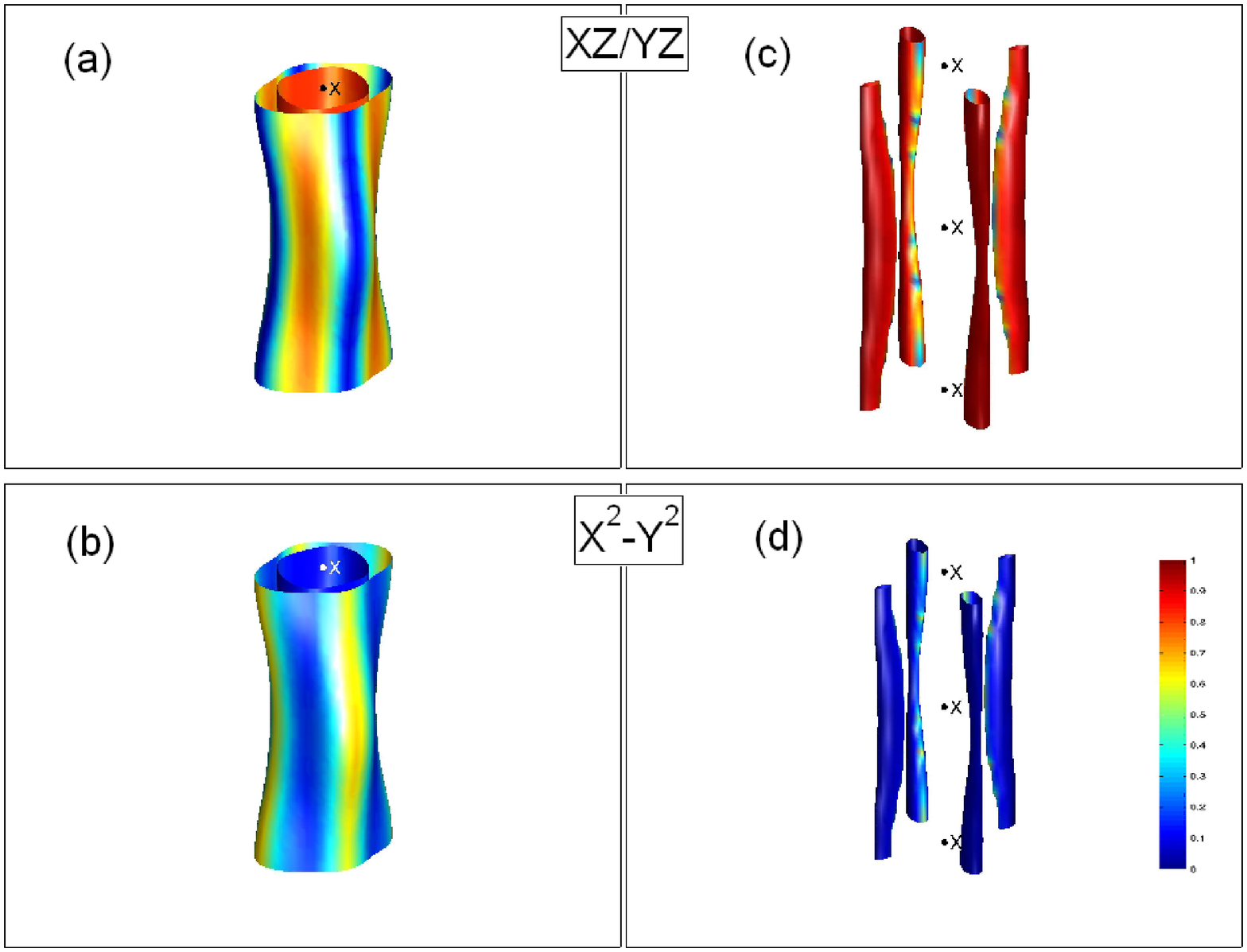}
\begin{flushleft}
{\label{FigS2} FIG.~S2: Three-dimensional Fermi surfaces of
KFe$_2$As$_2$ obtained from DFT calculations browsing the
contribution of $X^2-Y^2$ and $XZ$/$YZ$ orbitals to the Fermi
surfaces at X point. Panels a and b correspond to UD BaK obtained by
shifting $E_F$ towards higher positions.}
\end{flushleft}
\end{center}
\end{figure}

First we notice a difference in the FS topology between UD (a, b)
and KFe$_2$As$_2$ (c, d) cases where the FS at the X point has
changed from nearly-elliptical-like in the UD region to
propeller-like in the heavily OD region and KFe$_2$As$_2$. Moreover,
the dominant orbital characters on each FS sheet are browsed by the
corresponding colors. While both $X^2-Y^2$ and $XZ$/$YZ$ orbital
characters have considerable contributions at the X point in the UD
and nearly-OP regions (a, b), the $XZ$/$YZ$ orbital character
becomes dominant at that point in the OD region (c, d).

\newpage


\end{document}